\documentclass{article}

\usepackage{arxiv}

\usepackage[utf8]{inputenc} 
\usepackage[T1]{fontenc}    
\usepackage{hyperref}       
\usepackage{url}            
\usepackage{booktabs}       
\usepackage{amsmath}
\usepackage{amsfonts}       
\usepackage{nicefrac}       
\usepackage{microtype}      
\usepackage{lipsum}
\usepackage{color}
\usepackage{multirow}
\usepackage[normalem]{ulem}
\usepackage{graphicx}
\graphicspath{ {./images/} }

\title{Discovering associations in COVID-19 related research papers}

\author{
 Iztok Fister Jr. \\
  University of Maribor\\
  Maribor, Slovenia \\
  \texttt{iztok.fister1@um.si} \\
   \And
 Karin Fister \\
  Department of Infectious Diseases \\    
  General Hospital Murska Sobota\\
  Murska Sobota, Slovenia\\
  \And
 Iztok Fister \\
  University of Maribor\\
  Maribor, Slovenia \\
}

\begin{document}
\maketitle
\begin{abstract}
A COVID-19 pandemic has already proven itself to be a global challenge. It proves how vulnerable humanity can be. It has also mobilized researchers from different sciences and different countries in the search for a way to fight this potentially fatal disease. In line with this, our study analyses the abstracts of papers related to COVID-19 and coronavirus-related-research using association rule text mining in order to find the most interestingness words, on the one hand, and relationships between them on the other. Then, a method, called information cartography, was applied for extracting structured knowledge from a huge amount of association rules. On the basis of these methods, the purpose of our study was to show how researchers have responded in similar epidemic/pandemic situations throughout history. 
\end{abstract}


\section{Introduction}
When we look at the COVID-19 Global Cases web site~\cite{covid2020global,dong2020interactive} maintained by the Center for Systems Science and Engineering at Johns Hopkins University, we can observe with concern the comprehensiveness of the pandemic on the one hand, and an exponential increase in the number of infected people around the world on the other. While the conditions have stabilized in China, the circumstances in Europe and the USA have become critical. The number of infected people in Italy, the pandemic's epicenter in Europe, have achieved almost 60,000 at the time of this writing (i.e. on 23.3.2020), while the number of deaths is quickly approaching 7,000. 

Consequently, the mentioned circumstances have mobilized scientists from different domains around the world to try to find a way to throttle the coronavirus. These endeavors are not only the domain of researchers in medical labs, where they are searching for a new vaccine, but all the other mass of researchers from various scientific disciplines being indirectly affected. In this, data scientists also play an important role.

The present study associates the Association Rule Text Mining (ARTM)~\cite{fister2020population} method with information cartography~\cite{fister2020information}. The former is a data mining method used to search for interestingness terms and their mutual relations in the form of association rules. This method demands the parsing of text documents and highlights the words that are distinguished according to the appropriate measures. These words are called terms, while the relationships among them are described in the form of association rules. The latter is devoted to extracting structured knowledge from the huge amount of association rules generated in the first step. In line with this, the concept of information cartography has been applied~\cite{shahaf2015metro} which is capable of creating structured summaries of information, and visualize them in the form of metro maps. The role of the metro map in helping users  understand their surroundings, also is similar to the effect the metro map of information has on understanding information landscapes~\cite{shahaf2012trains}. Visualizations with metro maps can even tell stories to users and provide them with good directions. In essence, the metro maps consist of a set of metro lines, where each metro line interprets the same story from a different aspect. Metro stops on these lines introduce salient pieces of information (i.e. a definite term), while the interrelations among these pieces ensure the plot of the story. Recently, this methodology has been applied to understanding information in many areas~\cite{shahaf2012metro,shahaf2013information}. However, the concept of metro maps serves as a basis for exploring the extracted knowledge in this study.

The proposed method consists of the following steps: text preprocessing, generation of ARTM database, association rule simplification, word graph generation, metro map construction, and the exploration of extracted knowledge. In the first step, the interestingness words are extracted from a collection of observed paper abstracts. The association rules are generated from a set of mined words in the second step. The third step is devoted to simplification rules, where the rules with more antecedents and more consequent are simplified into a set of simple rules consisting of one antecedent and one consequent. These simple rules serve as building blocks for creating a word graph with source $X$ and sink nodes $Y$ connected with a directed arc, when there is an association rule $X\Rightarrow Y$ (fourth step). In the fifth step, the metro map is created from the word graph. Finally, the knowledge hidden in the metro maps is explored. 

The method was applied to a collection of paper abstracts found in the CORD-19 dataset~\cite{sebastian_kohlmeier_2020_3739581} in order to show how researchers have responded to similar epidemic/pandemic situations during history. Indeed, the results of the performed experiments have proven an increasing of terms referred to these situations.

In the remainder of the paper, the structure is as follows. Section~\ref{sec:2} introduces material and methods used in our study. In Section~\ref{sec:3}, the experiments are described and the obtained results are analyzed. A discussion of the results can be found in Section~\ref{sec:4}, while the paper is concluded with Section~\ref{sec:5}, which also provides an outline for the future work.

\section{Materials \& methods} \label{sec:2}

\subsection{Search Strategy}

The proposed method consists of three components: 
\begin{itemize}
\item ARTM:
\begin{itemize}
    \item text preprocessing, 
    \item generation of an ARTM database, 
\end{itemize}
\item information cartography:
\begin{itemize}
    \item association rule simplification, 
    \item term graph generation, 
    \item metro map construction, 
\end{itemize}
\item exploration of extracted knowledge.
\end{itemize}
The purpose of ARTM is to generate a database of the more interestingness terms. The information cartography enables us to extract the structured knowledge from a huge amount of association rules in the form of metro maps. The extracted knowledge in the form of terms, constituting the particular metro lines, serves as keywords for matching the terms from paper abstracts found in the huge database.

\subsubsection{ARTM}

\paragraph{Text preprocessing step.} Here, punctuation marks are removed as a first step. As a result, only words delimited by space remain in the document. Some words, like definite and indefinite articles (e.g. the, a, an), connective words (e.g. and, also, then), conjunctions (e.g. but, when, because), and verbs (e.g. is, done), represent so-called stop words, and must be removed next. The result of this removal is a sequence of terms. Then, the terms undergo term frequency calculation, where occurrences are not only determined, but also weighted. Here, a Term Frequency/Inverse Term Frequency (TF/ITF) weighting scheme is used that penalizes the rare occurring terms with higher weights.  

The TF/ITF weighting scheme is defined as follows: For the given term $z_j$, for $j=1,\ldots ,M$, occurring in document $d_i$, for $i=1,\ldots, N$, the term frequency is expressed as:
\begin{equation}
    \mathit{TF}_{i,j}=\frac{n(d_i,w_j)}{|d_i|},
\end{equation}
where $n(d_i,w_j)$ denotes the number of occurrences of term $w_j$ in document $d_i$, and $|d_i|$ is the total number of terms in document $D_i$. On the other hand, the inverse term frequency is expressed as:
\begin{equation}
    \mathit{ITF}_j=\left|\log \frac{n(d|w_j)}{N}\right|,
\end{equation}
where $n(d|w_j)$ denotes the number of documents $d$ containing the term $w_j$, where $N$ is the total number of documents.

Furthermore, the weighted frequency of the term $z_j$ in document $d_i$ is represented as a vector of weights $\mathbf{w}_i=\{w_{i,1},\ldots,w_{i,n}\}$, where each element $w_{i,j}$ is expressed as:
\begin{equation}
    w_{i,j}=\mathit{TF}_{i,j}\cdot \mathit{ITF}_j,\quad\text{for}~j=1,\ldots,n.
\end{equation}
Finally, the transaction database is generated from the relevant documents by moving each vector $\mathbf{w}_i$, representing weighted frequencies for all terms in the corresponding document, to a transaction database. In this way, the transaction database is very similar to the market basket, except that the weighted frequencies are put into transaction database instead of the value of one.

\paragraph{Generation of ARTM database step.} The ARTM problem is defined formally as follows: Let us assume a set of documents $D=\{d_1,\ldots,d_N\}$ and set of terms $Z=\{z_1,\ldots,z_M\}$, where $N$ denotes the maximum number of documents, and $M$ the maximum number of terms, respectively. Additionally, the matrix of weights $\mathbf{W}$ is assigned with the dimension $N\times M$, where each element $w_{i,j}$ represents a frequency weight of term $z_j$ in document $d_i$, calculated according a TF-ITF weighting scheme. Then, the task of generation is to select the binary vector $\mathbf{y}=(y_1,\ldots,y_M)^T$, determining the presence or absence of the corresponding term in the solution, such that the scalar product 
\begin{equation}
    \mathit{AWS}=\sum_{j=1}^{M}\sum_{i=1}^{N}{w_{i,j}\cdot y_j}
    \label{eq:aws}
\end{equation}
subject to
\begin{equation}
    \sum_{j=1}^M y_j\leq K,
    \label{eq:ineq}
\end{equation}
is maximum. Let us mention that variable $K$ denotes the maximum number of terms in an association rule. Actually, the selected elements of vector $\mathbf{y}$ form the set $Y=\{y_j|y_j=1,~\text{for}~j=1,\ldots,M\}$ that is a subset of $Z$, in other words $Y\subset Z$. Let us notice that the values of the vector are initially set to zero.  

Obviously, the problem is defined as an optimization and can be solved using any of the well-known stochastic population-based, nature-inspired algorithms. For our study, the Particle Swarm Optimization (PSO)~\cite{kennedy1995particle} was selected for this purpose. Interested readers, who would like to see the detailed implementation of this algorithm, are invited to consult the paper of Fister et al.~\cite{fister2020population}.

\subsubsection{Information cartography}

The concept of information cartography is applied in order to explore knowledge from an archive of mined association rules in text~\cite{fister2020information,shahaf2013information}, where this knowledge is visualized in the form of metro maps. The metro map is formally defined as $\mathcal{M}=(G,\Pi)$, where $G=(T,E)$ denotes a term graph of vertices $T=\{X_1,\ldots,X_N\}$, representing attributes, and edges $E=\{r_1,\ldots,r_M\}$, representing simple rules, together with the incident function $\psi_G$ that associates an ordered pair $\psi_G(r_k)=(X_i,Y_j)$ with direct edge $r_k$, when there exists a simple association rule in the form of $X_i\Rightarrow Y_j$, and $\Pi$ represents a set of paths in $G$. In the definitions, variables $N$ and $M$ denote the maximum number of vertices and maximum number of edges, respectively. 

\paragraph{Association rule simplification.} Thus, the simple association rule consists of only one antecedent and one consequent, where the former is mapped to the source node $X_i\in G$ and the latter to the sink node $Y_j\in G$ of the corresponding attribute graph, while the path $X_i\rightarrow Y_j$ leads from the source to the sink node.

In general, the association rules in the archive consist of more antecedents and more consequences, in other words:
\begin{equation}
    X_1 \wedge X_2 \wedge \ldots \wedge X_p \Rightarrow Y_1 \wedge Y_2 \wedge \ldots \wedge Y_q.
\end{equation}
The simple association rules are obtained from the mined rules by pairing each antecedent with each consequent, in other words:
\begin{equation}
    (X_1\Rightarrow Y_1),(X_1\Rightarrow Y_2),\ldots,(X_p\Rightarrow Y_q).
\end{equation}
In this process of simplifying rules, the $p\times q$ pairs of simple rules are obtained representing direct edges in the association graph.

\paragraph{Term graph generation.} The simple association rules present building blocks from which a term graph is constructed. In a term graph, each simple rule $X_i\Rightarrow Y_i$, for $i=1,\ldots,p\times q$, where $p$ designates the maximum number of antecedents and $q$ the maximum number of consequents, respectively, denotes a direct arc from source node $X_i$ to sink node $Y_i$. 

However, the nodes can appear in this graph as: (1) antecedent only, (2) consequent only, or (3) antecedent in one and consequent in the other rules. Consequently, these are divided into three subsets, i.e. $\mathit{Ante}(T)$, $\mathit{Cons}(T)$, and $\mathit{Mixed}(T)$. In the term graph $G$, the attributes in the antecedent subset $X\in \mathit{Ante}(A)$ represent source nodes with indegree zero, the attributes in consequent subset $Y\in \mathit{Cons}(A)$ are sink nodes with outdegree zero, while the attributes in the mixed subset $\langle X|Y\rangle\in \mathit{Mixed}(A)$ denote the intern nodes with an indegree and outdegree higher than zero. 

In summary, the antecedent set consists of nodes suitable for starting metro stops on the metro lines, the consequent set for the final metro stops, while the mixed set determines the intermediate metro stops and outlines a definite path towards achieving a certain final destination.

\paragraph{Metro map construction.} The task of metro map construction is to find a set of metro lines, where each metro line starts with the particular starting metro stop $X_i\in \mathit{Ante}(T)$ and finishes with the particular final metro stop $Y_i\in \mathit{Cons}(T)$, while the intermediate metro stops connect the starting metro stop with the final one by selecting proper simple rules from the term graph such that the sink node of the $i$-th simple rule is the source node of the $(i+1)$-th simple rule, in other words:
\begin{equation}
\underbrace{\underbrace{X_0\Rightarrow Y_0}_{\text{simple rule}~1}\equiv\underbrace{X_1\Rightarrow Y_1}_{\text{simple rule}~2}\equiv,\ldots,\equiv\underbrace{X_{n-1}\Rightarrow Y_{x-1}}_{\text{simple rule}~n}\equiv X_n}_{\text{metro line}}.
\label{eq:seq_rule}
\end{equation}
The terms $Y_i$ for $i=0,\ldots,(n-1)$ in Eq.~(\ref{eq:seq_rule}) can be avoided due to equivalence $Y_{i}\equiv X_{i+1}$. As a result, a sequence of implication rules is given, as follows: 
\begin{equation}
    X_{0}\Rightarrow X_{1}\Rightarrow, \ldots,\Rightarrow X_{n-1}\Rightarrow X_{n}.
    \label{eq:implic}
\end{equation}
According to standard rules in mathematical logic, Eq.~(\ref{eq:implic}) can be transformed, as follows:
\begin{equation}
X_{1}\wedge X_{2}\wedge\ldots\wedge X_{n-1}\Rightarrow X_{n},    
\label{eq:gold}
\end{equation}
asserting that the conjunctions of $(n-1)$ terms implied by the consequent is equivalent to a sequence of implications of $n$ term. Obviously, Eq.~(\ref{eq:gold}) is more easier to apply in an interpretation of the obtained results. 

The algorithm for constructing the metro map for visualizing the association rules needs to fulfill the following four objectives:
\begin{itemize}
    \item minimum line coherence,
    \item maximum map size,
    \item high coverage,
    \item high structure quality.
\end{itemize}
The minimum line coherence limits the number of intermediate metro stops in some metro line and is expressed by the following relation:
\begin{equation}
    \mathit{coherence}(\mathcal{M})\leq \tau,
    \label{eq:cons_1}
\end{equation}
where the variable $\tau$ determines the maximum number of intermediate metro stops. The maximum map size is referred to the maximum number of metro lines $L$, in other words: 
\begin{equation}
    |\mathcal{M}|\leq L.
    \label{eq:cons_2}
\end{equation}
Indeed, we are interested in covering our information domain by using the number of metro lines as close to $K$ as possible.

The coverage estimates how well the selected metro line exploits the attributes in a transaction database. In line with this, the lift measure of association rule $\mathit{Lift}(X\Rightarrow Y)$ is used that is expressed as:
\begin{equation}
    \mathit{Lift}(X\Rightarrow Y)=\frac{\mathit{supp}(X\cup Y)}{\mathit{supp}(X)\times\mathit{supp}(Y)}.    
\end{equation}
Let it be noted that the characteristic of the measure is that the higher the value, the stronger the association. Additionally, the coverage of the whole metro line $\pi \in \Pi$ is expressed as:
\begin{equation}
    \mathit{coverage}(\pi)=\frac{1}{|\pi|}\sum_{r\in\pi}{\mathit{Lift}(r)},
\end{equation}
where $r$ represents the particular simple association rule $X\Rightarrow Y$. Finally, the coverage of the metro map is a simple average of all the proposed metro lines, in other words:
\begin{equation}
    \mathit{coverage}(\Pi)=\frac{1}{|\mathcal{M}|}\sum_{\pi\in\Pi}{\mathit{coverage}(\pi)}.
    \label{eq:cov}
\end{equation}

The metro map structure quality refers to the diversity of the metro lines, where we are interested in those metro lines that differ in the intermediate points as much as possible. This relation is expressed by the following equation:
\small
\begin{equation}
    \mathit{sQuality}(\mathcal{M})=\frac{1}{C}\sum_{\substack{\pi_i\in\Pi \\ \pi_j\in\Pi\\i\neq j}}{\frac{|\{r\in\pi_i\wedge s\in\pi_j:r\neq s\}|}{|\pi_i|\times|\pi_j|}},
    \label{eq:qual}
\end{equation}
\normalsize
where the variable $C=\binom{|\mathcal{M}|}{2}$ counts the number of metro line interactions.

In summary, the quality of the solution considers the constructed metro map according to two objectives: the coverage (according to Eq.~(\ref{eq:cov})), and the quality (according to Eq.~(\ref{eq:qual})). Both equations are contained within a linear combination as follows:
\begin{equation}
    f(\mathbf{y}_i)=(\mathit{coverage}(\mathbf{y}_{i})+w\cdot(1-\mathit{sQuality}(\mathbf{y}_{i})))\cdot n_i,
    \label{eq:fit}
\end{equation}
where the weight variable $w$ indicates the influence of the second term on the total fitness value, and $n_i$ is the number of metro lines. However, each solution is subject to minimum line coherence, and maximum map size as previously explained.

A stochastic population-based, nature-inspired evolutionary algorithm was used for the implementation of the metro map construction. Interested readers are invited to consult the paper of Fister et al.~\cite{fister2020information} for more details about the implementation of the evolutionary algorithm.

\subsubsection{Exploration of extracted knowledge}
Normally, the created metro map of ARTM information is visualized in the sense of real metro maps, where each metro line consists of a particular number of metro stops. Some metro lines proceed straightforwardly, while some can interrelate between the other. Obviously, these relationships affect the plot of the story and highlight special events that can occur either unexpectedly or as an ordinary consequence of some process operation. 

In our study, we are interested in identifying the terms that occur in the best metro map according to the fitness function evaluation. These terms, then, serve as keywords for searching for knowledge hidden in articles of papers saved into a huge database. The results of these experiments are then visualized using traditional statistical visualization techniques.

\subsection{Parameter setting}
The proposed search method introduces two stochastic, nature-inspired, population-based algorithms: The former searches for the optimal binary vector $\mathbf{y}$, in which the value 0 determines that the corresponding term is absent from the solution and value 1 that it is present in the solution. The latter is reserved for constructing the metro map. Both algorithms are controlled by some parameters that ensure their proper operation. The parameter setting used during the experimental work is illustrated in Table~\ref{tab:1}.
\begin{table}[htb]
    \caption{Parameter setting of evolutionary algorithm for metro map creation.}
    \label{tab:1}
    \centering
    \begin{tabular}{l|c|r}
        Parameter & Abbreviation & Value \\ \hline
        Maximum number of terms in association rules & $K$ & 10 \\
        Maximum number of intermediate metro stops & $\tau$ & 10 \\
        Maximum number of metro lines & $L$ & 10 \\
        Weight variable & $w$ & 0.5 \\
    \hline     
    \end{tabular}
\end{table}

Let us mention that the problem parameters are only captured in the table. The proper values of these parameters were found after extensive experimental work. Indeed, the detailed setting of algorithm parameters can be found by interested readers in the corresponding literature~\cite{fister2020population,fister2020information}.

\subsection{Data Extraction and Outcome Measure}
The study was divided into two parts: In the first part, the ARTM was conducted on the CORD-19 dataset~\cite{sebastian_kohlmeier_2020_3739581}\footnote{https://pages.semanticscholar.org/coronavirus-research}, where its non-commercial subset was taken into consideration, while the quality of solutions were evaluated by maximizing Eq.~(\ref{eq:aws}). The second part was applied on the MEDLINE\footnote{ftp://ftp.ncbi.nlm.nih.gov/pubmed/baseline/} database. This database consists of medical scientific papers. Here, the abstracts of all the papers found in the database were parsed using the tool Pubmed Parser in Python~\cite{achakulvisut2020pubmed}. In this case, the quality of solutions are estimated using Eq.~(\ref{eq:fit}).

\section{Results} \label{sec:3}
\begin{figure}[htb]
  \centering
    \includegraphics[width=1.0\textwidth]{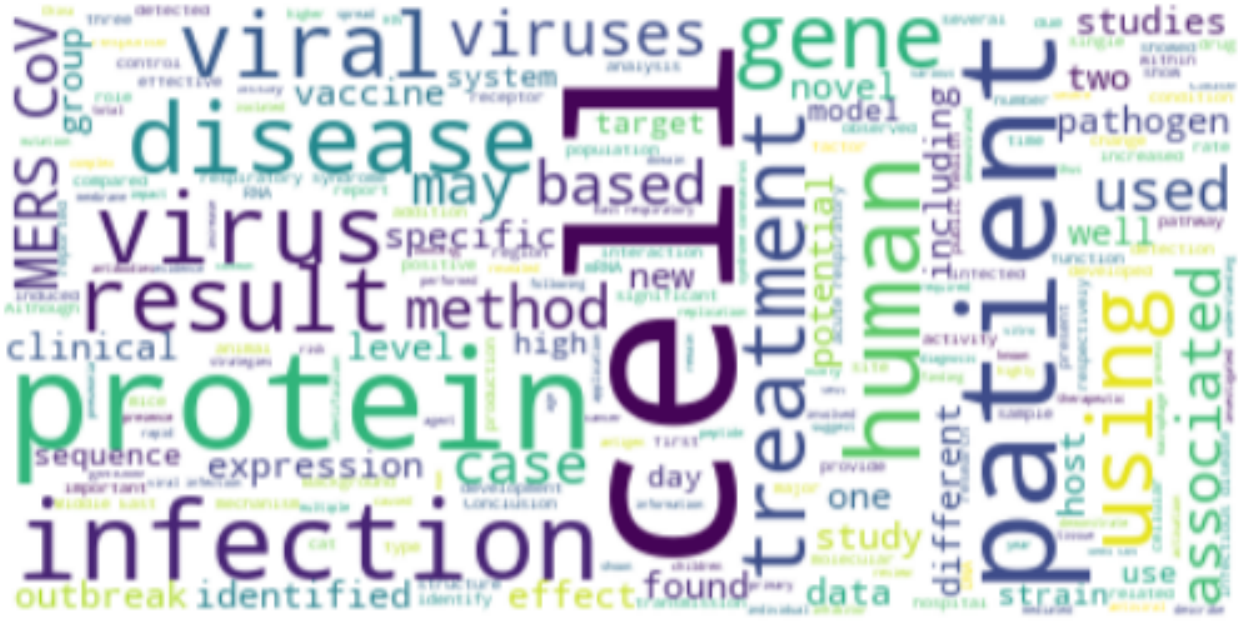}
 \caption{A word cloud.}
 \label{fig:cloud}
\end{figure}

The purpose of our experimental work was to show how researchers have responded to similar epidemic/pandemic situations throughout history. In line with this, the ARTM method was applied to the CORD-19 dataset. The results of the method is presented in the word cloud in Fig.~\ref{fig:cloud}, 
from which it can be seen that terms like ''cell'', ''protein'', ''infection'', and ''patient'' occur most frequently in the observed abstracts of the papers. 

The goal of our research was to show how researchers reacted to epidemic/pandemic events in the past. In line with this, terms in association rules constituting the best metro maps according to the fitness function were extracted (i.e., 44 such terms), from which those terms that do not have any connections with medicine were eliminated (i.e., 18 terms). Finally, the 26 terms that remained are illustrated in Table~\ref{tab:1}.
\begin{table}[htb]
    \caption{The best metro map found using the evolutionary algorithm.}
    \label{tab:1}
    \centering
    \begin{tabular}{l|l}
        Num. & Metro line description  \\ \hline
        1 & \sout{via}$\Rightarrow$mitochondrial \\
        2 & \sout{produced}$\Rightarrow$propagation \\
        3 & \sout{three}$\wedge$\sout{study}$\wedge$transfection$\wedge$human$\wedge$\sout{severe}$\wedge$viruses$\wedge$pneumonia$\Rightarrow$ventilation \\
        4 & \sout{act}$\Rightarrow$pseudoknots \\
        5 & \sout{consistent}$\wedge$rna$\wedge$virus$\wedge$transfection$\wedge$viral$\wedge$\sout{review}$\wedge$\sout{identify}$\wedge$\sout{study}$\wedge$\sout{caused}$\Rightarrow$cardiac \\
        6 & \sout{introduced}$\Rightarrow$diagnostic \\
        7 & quarantine$\Rightarrow$\sout{taking} \\
        8 & pathogens$\wedge$diseases$\Rightarrow$\sout{truncated} \\
        9 & \sout{like}$\wedge$transfection$\wedge$\sout{different}$\wedge$virus$\wedge$protein$\wedge$cells$\wedge$\sout{study}$\wedge$viruses$\wedge$human$\wedge$\sout{may}$\Rightarrow$downregulation \\
        10 & pulmonary$\Rightarrow$h7n9 \\
    \hline     
    \end{tabular}
\end{table}

\begin{figure}[htb]
  \centering
    \includegraphics[width=1.0\textwidth]{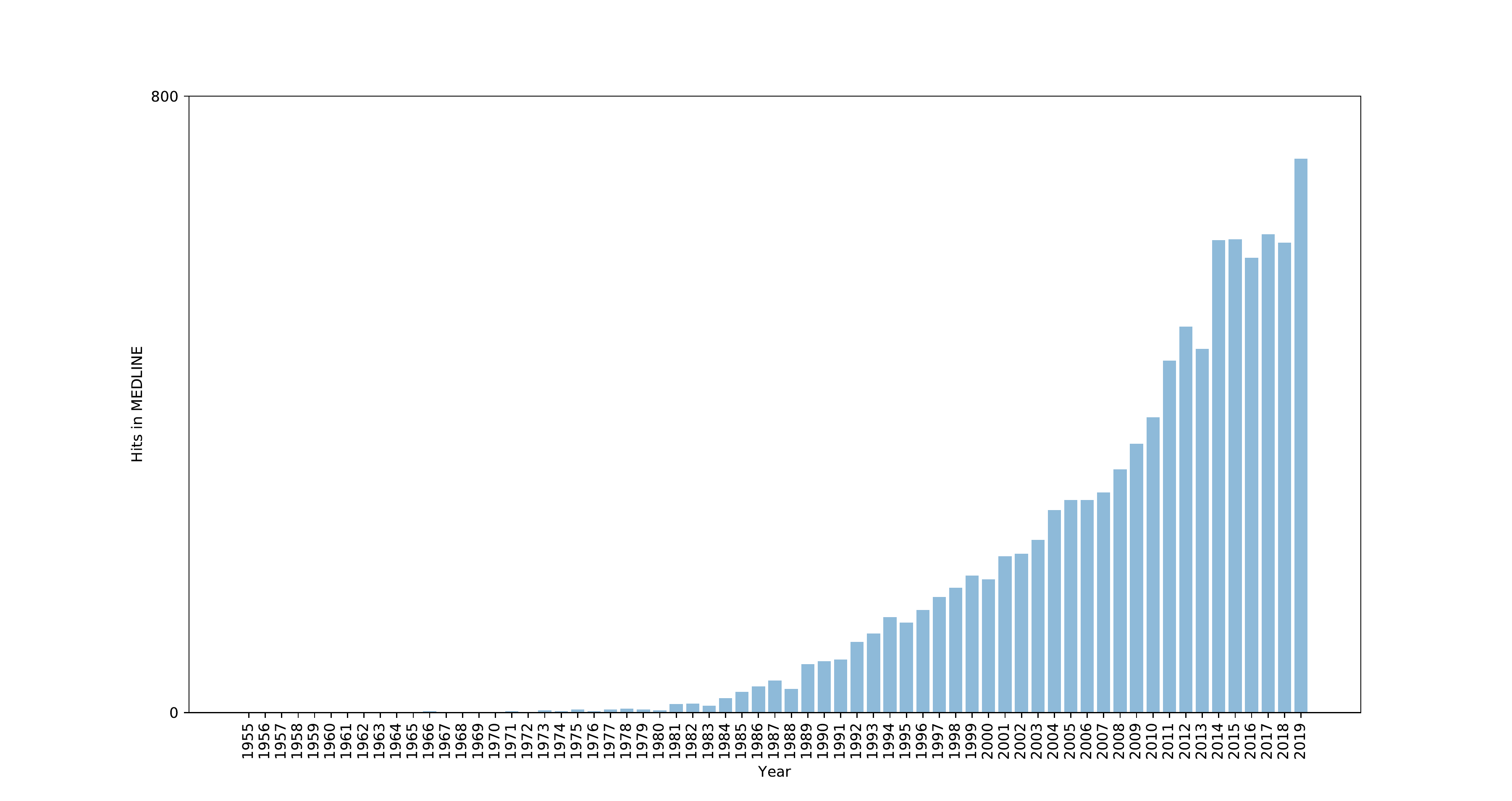}
 \caption{Historical review of mined terms according to Table~\ref{tab:1}.}
 \label{fig:hist}
\end{figure}

Let us mention that the eliminated terms are denoted as crossed out text in the table. All the other regular terms (i.e. 21 without any repetition of the same words) are entered into the second phase of the experiment, where they were used as keywords to searching for the abstracts of medicine papers maintained in the MEDLINE database from the year 1955 onward. All abstracts matching at least 30~\% of the keywords contribute to the final outcome. The number of hits are depicted in Fig.~\ref{fig:hist}.

Interestingly, the number of hits increased from the year 1955 until 2019 almost exponentially, although there are some periods of stall (e.g., the year 2013, or period from years 2014 to 2018). This means that the application of terms like ''viruses'', ''quarantine'', and ''h7n9'' have appeared with greater frequency in line with the appearance of different viruses denoting epidemic/pandemic events over recent years. Once again, the increasing can be observed in the year 2019.

\section{Discussion} \label{sec:4}
The results of the study showed that epidemic/pandemic events affected the production of new scientific papers a lot during the course of history. On the other hand, these were inspired by the  emergence of new viruses or caused a mutation of old ones. A historical analysis revealed the biggest increase in the number of papers in the years 2013 and 2014. This increase correlates with the outbreak of the MERS disease. Interestingly, in these times, preprints were not as popular as today~\cite{kaiser2017preprints}. Therefore, a lot of papers struggled during the long-term review process and appeared many months after the outbreak. Among the terms found using the proposed method, the papers that referred to any aspect correlated with coronavirus research were mostly distinguished by the types of the virus (e.g. RNA), its clinical manifestation (e.g. pneumonia), and consequences (e.g. quarantine), the familiarity (e.g. H7N9), or virus description (e.g. pathogen). Some terms found in the study were hard to define like mitochondrial and pseudoknots. 

\section{Conclusion} \label{sec:5}
The COVID-19 pandemic has affected the lives of people all over the world. Social isolation and quarantines stopped the world for many months. Moreover, the catastrophe has entered its zenith at this time. Although no one in the world expected such dimensions for the pandemic, the situation has shown how susceptible humanity can be. 

The purpose of the study was to show how researchers responded with subjects of their papers in similar epidemic/pandemic situations during history. This study analyzed the abstract of the papers found in the CORD-19 dataset using the ARTM method and extracted knowledge hidden in a large amount of mined association rules with metro map methodology. The extracted terms were then used as keywords to search the abstracts of papers collected in the MEDLINE database. 

The results of the study showed that the number of papers that include the terms proposed by the metro map method increased exponentially during the course of history. In the future work, we will try to relate these findings to the increased usage of antiviral drugs. We speculate that higher consumption of antiviral drugs may lead to the development of more pathogenic strains like SARS-CoV-2.


\begin{thebibliography}{10}

\bibitem{achakulvisut2020pubmed}
Titipat Achakulvisut, Daniel Acuna, and Konrad Kording.
\newblock Pubmed parser: A python parser for pubmed open-access xml subset and
  medline xml dataset xml dataset.
\newblock {\em Journal of Open Source Software}, 5(46):1979, 2020.

\bibitem{dong2020interactive}
Ensheng Dong, Hongru Du, and Lauren Gardner.
\newblock An interactive web-based dashboard to track covid-19 in real time.
\newblock {\em The Lancet infectious diseases}, 2020.

\bibitem{fister2020population}
Iztok Fister~Jr, Suash Deb, and Iztok Fister.
\newblock Population-based metaheuristics for association rule text mining.
\newblock {\em arXiv preprint arXiv:2001.06517}, 2020.

\bibitem{fister2020information}
Iztok Fister~Jr and Iztok Fister.
\newblock Information cartography in association rule mining.
\newblock {\em arXiv preprint arXiv:2003.00348}, 2020.

\bibitem{covid2020global}
Center for Systems~Science and Engineering.
\newblock Coronavirus {COVID-19} {G}lobal {C}ases,
  https://coronavirus.jhu.edu/map.html, 2020.

\bibitem{kaiser2017preprints}
J~Kaiser.
\newblock Are preprints the future of biology? a survival guide for scientists.
\newblock {\em Science}, 485, 2017.

\bibitem{kennedy1995particle}
J~Kennedy and R~Eberhart.
\newblock {Particle swarm optimization}.
\newblock In {\em Neural Networks, 1995. Proceedings., IEEE International
  Conference on}, volume~4, pages 1942--1948. IEEE, 1995.

\bibitem{sebastian_kohlmeier_2020_3739581}
Sebastian Kohlmeier, Kyle Lo, Lucy~Lu Wang, and JJ~Yang.
\newblock Covid-19 open research dataset (cord-19), March 2020.

\bibitem{shahaf2012metro}
Dafna Shahaf, Carlos Guestrin, and Eric Horvitz.
\newblock Metro maps of science.
\newblock In Qiang Yang, Deepak Agarwal, and Jian Pei, editors, {\em KDD},
  pages 1122--1130. ACM, 2012.

\bibitem{shahaf2012trains}
Dafna Shahaf, Carlos Guestrin, and Eric Horvitz.
\newblock Trains of thought: Generating information maps.
\newblock In {\em Proceedings of the 21st International Conference on World
  Wide Web}, WWW ’12, page 899–908, New York, NY, USA, 2012. Association
  for Computing Machinery.

\bibitem{shahaf2015metro}
Dafna Shahaf, Carlos Guestrin, Eric Horvitz, and Jure Leskovec.
\newblock A metro map can tell a story, as well as provide good directions.
\newblock {\em Communications of the ACM}, 58(11):62--73, November 2015.

\bibitem{shahaf2013information}
Dafna Shahaf, Jaewon Yang, Caroline Suen, Jeff Jacobs, Heidi Wang, and Jure
  Leskovec.
\newblock Information cartography: Creating zoomable, large-scale maps of
  information.
\newblock In {\em Proceedings of the 19th ACM SIGKDD International Conference
  on Knowledge Discovery and Data Mining}, KDD ’13, page 1097–1105, New
  York, NY, USA, 2013. Association for Computing Machinery.

\end{thebibliography}
\end{document}